\documentclass[aip,amsmath,amssymb,preprint]{revtex4-1}

\usepackage{soul}
\usepackage{color}
\usepackage{graphicx}

\usepackage[normalem]{ulem}
\usepackage{CJK}
\usepackage{threeparttable}

\newcommand{\stmod}[1]{{\color{black} #1}}
\newcommand{\stcomt}[1]{{\color{black} #1}}
\newcommand{\paddyspeaks}[1]{{\color{black} #1}}

\begin{document}

\begin{CJK*}{UTF8}{} 
\CJKfamily{min}

\title{Towards the ideal glass transition by pinning in a dimer-polymer mixture}
\author{Genki Kikumoto (菊本 元気)}
\affiliation{Kyoto Institute of Technology, Hashiue-cho, Matsugasaki, Kyoto, 606-8585, Japan}
\author{Naohiro Torii (鳥居 直弘)}
\affiliation{Kyoto Institute of Technology, Hashiue-cho, Matsugasaki, Kyoto, 606-8585, Japan}
\author{Koji Fukao (深尾 浩次)}
\affiliation{Ristumeikan University, 1-1-1, Noji-Higashi, Kusatsu-shi, Shiga, 525-8577, Japan}
\author{C. Patrick Royall}
\affiliation{Gulliver UMR CNRS 7083, ESPCI Paris, Universit\' e PSL, 75005 Paris, France.}
\affiliation{H.H. Wills Physics Laboratory, Tyndall Avenue, Bristol, BS8 1TL, UK}
\affiliation{School of Chemistry, University of Bristol, Cantock's Close, Bristol, BS8 1TS, UK}
\affiliation{Centre for Nanoscience and Quantum Information, Tyndall Avenue, Bristol, BS8 1FD, UK}
\author{Haruhiko Yao (八尾 晴彦)}
\affiliation{Kyoto Institute of Technology, Hashiue-cho, Matsugasaki, Kyoto, 606-8585, Japan}
\author{Yasuo Saruyama (猿山 靖夫)}
\affiliation{Kyoto Institute of Technology, Hashiue-cho, Matsugasaki, Kyoto, 606-8585, Japan}
\author{Soichi Tatsumi (辰巳 創一)}
\affiliation{Kyoto Institute of Technology, Hashiue-cho, Matsugasaki, Kyoto, 606-8585, Japan}\date{\today}

\begin{abstract}
We use a mixture of a polymer and its dimer to control dynamics in a manner inspired by \emph{pinning} a fraction of the system. In our system of $\alpha$-methyl styrene, where the polymer has a glass transition at higher temperature than the dimer, at intermediate temperatures, the polymer acts to ``pin'' the dimer. Within this temperature range, we use differential scanning calorimetry to infer a point-to-set length which we find to be profoundly influenced by the degree of pinning. We determine the dynamics of the system with dielectric spectroscopy and find that while the dynamics are very substantially slowed by the ``pinning'', the fragility exhibits only a small change
\paddyspeaks{relative to the precision of our measurements.} This may indicate that \paddyspeaks{in the approach we have used,} fragility has \paddyspeaks{a relatively weak} 
dependence on quantities such as the point--to--set length. 
\paddyspeaks{However} 
an alternative explanation is that the dimer may act to \emph{plasticize} the polymer and thus open routes to relaxation that may be inaccessible to fully pinned systems.
\end{abstract}

\pacs{}

\maketitle
\end{CJK*}

\section{Introduction}
The mechanism by which the viscosity of liquids increases by more than 14 orders of magnitude over a small change in temperature or density, the glass transition, remains one of the major challenges of condensed matter physics \cite{berthier2011}. One enduring idea of the origins of such a massive increase in relaxation timescales is the drop in entropy in the supercooled liquid \cite{royall2018, berthier2019arxiv}, which may be related to a so-called ``ideal glass'', an amorphous state with sub-extensive configurational entropy encountered at a temperature $T_\mathrm{K}$. Such a system should exhibit a divergent static lengthscale corresponding to the \emph{amorphous order} related to the drop in configurational entropy. Because the system falls out of equilibrium at the operational glass transition $T_\mathrm{g}>T_\mathrm{K}$, this putative state is remarkably hard to access in experiment or computer simulation, \paddyspeaks{despite recent developments \cite{swallen2007,kearns2008,ninarello2017,royall2018,ortlieb2021}.}

\begin{figure}
\begin{center}
\includegraphics[width=0.45\textwidth]{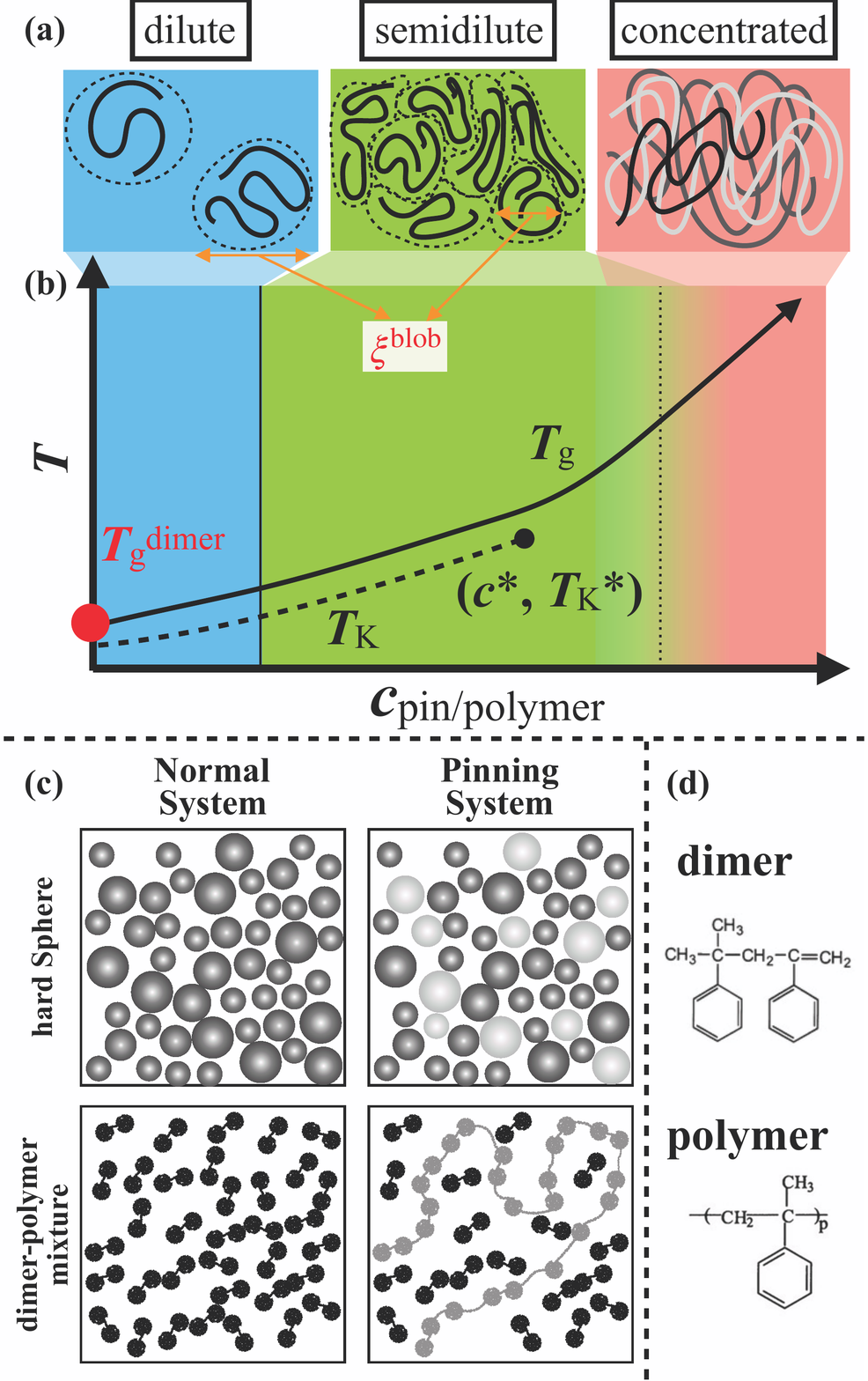}
\caption{
(color on line) 
(a) Schematics of the dilute (blue), semidilute (light green), and concentrated regimes (pink). The boundary of the semidilute and concentrated regimes is empirically determined and hence shaded whereas the boundary of dilute and semidilute regimes is precisely defined \cite{teraoka}. $\xi^\mathrm{blob}$ indicates the typical size of blobs in polymers with corresponding polymer concentration which is described in Eq.\ref{eqBlobsize}. 
(b) Schematic of the pinning phase diagram \cite{cammarota2012pnas}. Thick line and dotted line indicate $T_\mathrm{g}$ and $T_\mathrm{K}$, respectively. $(c^*, T_\mathrm{K}^*)$ denotes the critical point where the ideal glass vanishes. Regimes are schematically divided by dashed lines into dilute (blue), semidilute (light green), and concentrated regimes (pink) as shown in the figure. 
(c) Schematic view of pinning for supercooled liquid for e.g. hard spheres (upper panels) and 
 ``pinning'' in the dimer polymer mixture (lower panels), respectively. While immobilized particles in the supercooled liquid (from upper left to upper right) are pinned particles. Here we introduced the polymer into the pure dimer liquid (from lower left to lower right) which plays a role as ``pinned'' particles due to the low mobility of the chain. The pinning fraction in both systems corresponds to the ratio of immobilizing particles and the polymer concentration, respectively. In those panels, pinned particles are illustrated in grey compared with normal molecules. 
(d) Chemical structures of the dimer (2,4-diphenyl-4-methyl-1-pentene) and polymer, respectively. 
}\label{figSchematic} 
\end{center}
\end{figure}

A major development in approaching the ideal glass is \emph{pinning} \cite{biroli2008,cammarota2012pnas}, where a subset of the system is frozen. Effectively, when the static length scale of amorphous order is comparable to the separation of the pinned particles, the system can undergo an arrest reminiscent of an ideal glass transition. Crucially, this occurs at higher temperature where the dynamics of the unpinned system is amenable to computer simulation \cite{berthier2012,kob2013,ozawa2015}, or experiments with colloids \cite{ganapathy2014,williams2015}. Intriguingly, the smaller dynamic correlation length resulting from pinning was related to a decrease in fragility in computer simulation \cite{chakrabarty2015}.

Despite the strengths of the pinning concept, direct application to experiments on molecular liquids remains very challenging, since it is necessary to equilibrate the system and then somehow immobilize a subset $c$ of pinned particles. However, it is possible to investigate a closely related phenomenon, by mixing the system with a second, larger, species which exhibits much slower dynamics, so--called soft pinning  \cite{das2017,das2021}. Alternatively, one may use an immobilized subset of the system by taking the \emph{polymer} of the constituent molecules. This is the approach that we pursue here. In particular we carry out thermodynamical studies by using differential scanning calorimetry (DSC) to investigate the glass transition in a dimer-polymer mixture in which the polymer has a glass transition $T_\mathrm{g}$  at a rather higher temperature than that of the dimer. Vitrification of the polymer then acts in a manner similar to pinning a subset of the system.

From analysis of our differential scanning calorimetry data, we infer an abrupt change in size of the dynamic correlation length at the glass transition temperature as a function of pinning fraction. In particular, for weaker pinning ($\lesssim$20 wt\%), we find a constant co-operativity length. In the stronger pinning regime ($\gtrsim$20 wt\%), we find a drop in the cooperativity length with increasing polymer concentration. We further explore the dynamics of our pinned system using dielectric spectroscopy. 
\paddyspeaks{In particular, we obtain a measure of the structural relaxation time as a function both of temperature and of the polymer concentration. In the resulting ``Angell'' plots, fragility is weakly influenced by the ``pinning'' of the polymer.}

\paddyspeaks{This paper is organised as follows. In Section \ref{sectionStrategy}, we outline our strategy to approach pinning--like behaviour in a molecular system. We follow with a description of our experimental methods in Section \ref{sectionExperimental}.  In our results section \ref{sectionResults}, we begin by explaining the approach we take to determine the glass transition in our dimer--polymer system (Section \ref{sectionElucidation}) before presenting our results from differential scanning calorimetry in section Section \ref{sectionChange}. The results of our dielectric spectroscopy measurements are shown in Section \ref{sectionDielectric}. We discuss and conclude our findings in Section \ref{sectionDiscussion}.}

\section{Strategy to approach pinning in experiment}
\label{sectionStrategy}

Polymers have a range of concentration \emph{regimes}, and here we are concerned with the semi-dilute and concentrated regimes. In our system, the crossover between these occurs at a polymer concentration around 30 wt\%. In the semidilute regime, there is sufficient space between polymer chains for the dimer molecules to re-arrange. A schematic of these regimes is shown in Fig. \ref{figSchematic}(a).

Traditionally, the increase of mobility induced by the addition of small molecules is termed plasticization. Here, pinning a small molecule glassformer via the introduction of its polymer is a related phenomenon, but the dynamics are viewed from the perspective of the monomer rather than the polymer. For example, in the case of the toluene-polystyrene mixture, where the polymer has a similar chemical structure to the toluene, a separation of relaxation timescales involving the appearance of two glass transition temperatures is found \cite{adachi1975, floudas1993}.

For the system of interest here, the dimer-polymer mixture of $\alpha$-methyl styrene,  the pentamer-polymer and hexamer-polymer mixtures have been investigated \cite{huang2003, zheng2008}, mainly in the concentrated regime, (higher than $\sim$30 wt\% polymer concentration) where there is insufficient space between polymer segments for the molecular and oligomeric units to re-arrange between the (arrested) polymer chains, and so they do not dominate the dynamics of the system. Here, on the other hand we consider a larger range of polymer concentration, $c^\mathrm{pol}$ 1--50 wt\%, including the semidilute regime where there is \paddyspeaks{sufficient} 
space between the polymer chains for the dimers to re-arrange. \paddyspeaks{In this case, the dimer} 
should dominate the relaxation dynamics.

\section{Experimental}
\label{sectionExperimental}

\subsection{Sample preparation}
We used the $\alpha$-methyl styrene dimer (2,4-diphenyl-4-methyl-1-pentene) and its polymer. The dimer was obtained from TCI Co.,Ltd (Tokyo, Japan) with purity of above 95\%. The polymer was obtained from Polymer Source, Inc. (Dorval, Canada). The molecular weight and polydispersity of polymer were $M_{\rm w}= 3.74\times 10^5$ and $M_{\rm w}/M_{\rm n}= 1.10$, respectively. The pure dimer and polymer have glass transitions ($T_\mathrm{g}$, \textit{ie} structural relaxation time of 100s) of 193 K and 420 K respectively. Therefore, in these semi-dilute polymer solutions, we presume the polymers are ``immobilized'' compared to the dimers for the temperature range $193$ K $< T < 420$ K. The mixture of dimer and polymer was prepared at polymer concentrations $c^\mathrm{pol}$ of 1 to 50 wt\%. The dimer and polymer were codissolved in toluene, which was subsequently removed by heating to 50 $^\circ$C. Residual quantities of toluene and the polymer concentration were determined by with $^1$H nuclear magnetic resonance.

\subsection{Heat capacity determination}
A Shimadzu DSC-60 differential scanning calorimeter with a nitrogen coolant system was used to obtain the heat capacities of the samples. All measurements were made in a nitrogen atmosphere. The sample mass was varied from 4.15 to 11.31 mg, and the thickness was less than 0.5 mm. Samples were initially equilibrated at room temperature, which is sufficiently larger than the glass transition temperature of the dimers, and then, quenched at $10 \mathrm{K} / \mathrm{min}$ to $130 \mathrm{K}$, and finally equilibrated at 140 K for 20 minutes to begin heat capacity measurements. For all samples, heating--cooling cycles from 140 K to 380 K were applied twice to confirm reproducibility.
To avoid the effect of overshooting heat capacities during a heating scan, further analysis was employed from the cooling scan.

\subsection{Dielectric relaxation spectroscopy}
Dielectric relaxation (DR) measurements were performed using an impedance analyzer(Beta analyzer, Novocontrol Technologies, Montabaur, Germany) with a frequency $\omega$ range from $6.28\times 10^{-2}$ to 9.42 $\times 10^{6}$ $\mathrm{s}^{-1}$, in which the temperature was changed from 353 K to 163 K with a typical temperature step of 10 K, using a temperature control system (the Quatro Cryosystem, Novocontrol Technologies). Samples were measured under isothermal condition after equilibration where temperature variations were less than 1 K. Samples were filled in liquid parallel plate sample cells (BDS1308) which have 50 $\mathrm{\mu m}$ gaps between electrodes at room temperature, where samples were in liquid state. The diameter of the circular electrodes was 20 mm. The details of the measurement system may be found in the literature \cite{tahara2010, taniguchi2015}.

\begin{figure}
\begin{center}
\includegraphics[width=0.4\textwidth]{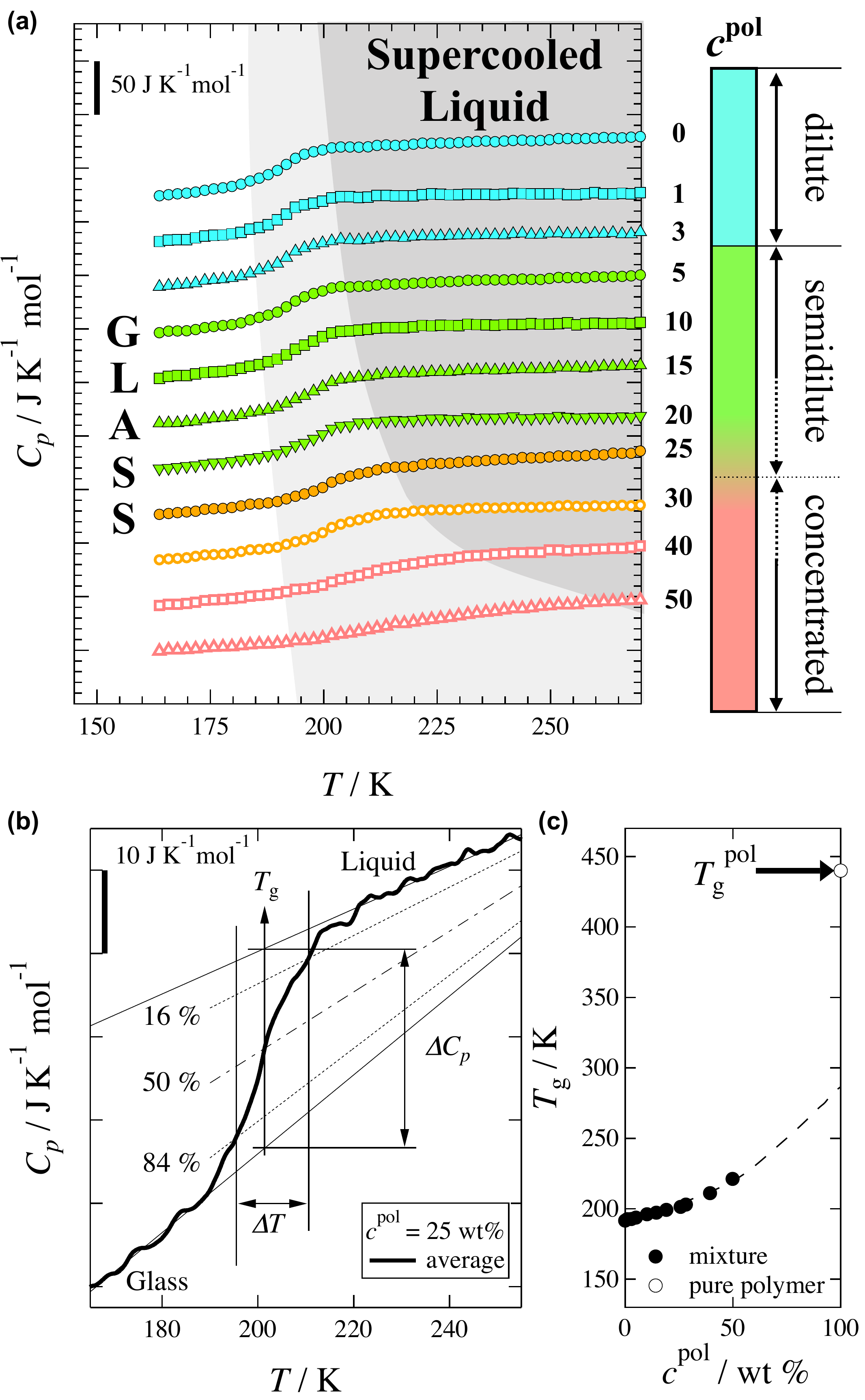}
\caption{(color online) Using pinning to change the glass transition in the dimer-polymer mixture of $\alpha$-methyl styrene.
(a) Heat capacities obtained from DSC measurements. Curves are shifted for clarity. Data of $c^\mathrm{pol}$ from 0 to 50 wt\% are shown. Each curve is averaged by using two independent scans. 
The left- and right hand axes indicate molar isobaric heat capacities per dimer and weight concentration of polymer, respectively. 
Unshaded and dark shaded regions indicate glass and liquid and light shaded regions indicate the temperature range of the glass transition.
(b) Full scanning heat capacities in 25 wt\% polymer concentration and schematics of glass transition analysis. 
Thick curve: averaged heat capacity. Thin lines: linear extrapolations of glassy and liquid heat capacities. Dotted lines: 16:84 and 84:16 mixtures of glassy and liquid heat capacities. Chain line: 50:50 mixture of glass and liquid heat capacities. 
(c) $T_{\mathrm g}$s (filled circles) obtained as a function of $c^{\mathrm pol}$ and of pure polymer (open circles, indicated by arrow). Dotted line is a guide to the eye to show an extrapolation of $T_{\mathrm g}$s of the mixture. 
\label{figCp}}
\end{center}
\end{figure}

\section{Results}
\label{sectionResults}

\subsection{Identifying the glass transition in the \stmod{dimer-polymer}
system}
\label{sectionIdentifying}
As shown in Fig.~\ref{figCp}(a), the glass transition \paddyspeaks{temperature} was detected as steps of heat capacities. Upon increasing the polymer weight fraction $c^\mathrm{pol}$, $T_{\mathrm g}$ increases and the ``transition'' broadens. However, comparing the difference in $T_{\mathrm g}$ between the \paddyspeaks{pure dimer and the pure polymer}, which is about \paddyspeaks{220} K, the increase in $T_{\mathrm g}$ is limited to about 30 K even for a polymer concentration of 50 wt\%, as shown in Figure \ref{figCp}(c).



\paddyspeaks{Consistent with}
our results, previous work on low molecular weight materials-polymer mixtures where the system is in the semidilute regime, found the increase in $T_{\mathrm g}$ from the pure dimer to be very small ($<10$ K). Those dependencies of $T_{\mathrm g}$ upon composition are not explained by traditional models \cite{adachi1975, scandola1982, pizzoli1983, scandola1987, floudas1993}, which assume a smooth functional form upon composition \cite{gordon1952,fox1956,kelley1961,kwei1984,lodge2000}; we note these models mostly consider a binary mixture of different polymers. In fact, Scanodola {\it et al.} reported a deviation from a traditional model especially in the semidilute regime for a mixture of polymer and low molecular weight liquid, polyvinylchloride and dimethylphthalate \cite{scandola1982,pizzoli1983}. 



\subsection{Elucidation of the pinning length scale}
\label{sectionElucidation}

We used Donth theory \cite{donth1996,sillescu1999} to determine the number of \emph{dimers} ($N^{(\alpha)}$) or characteristic volume of cooperative motion associated with alpha relaxation which corresponds to cooperative motion in the liquid. This separates the system into a large number of volumes ($V^{(\alpha)}$), \emph{which are assumed to be non-interacting}. Each volume is taken to have its own glass transition temperature. The energy is further assumed to have vibrational contributions $E^\mathrm{(vib)}$ and contributions due to alpha relaxation, $E^{(\alpha)}$, the latter is assumed not to occur below $T_\mathrm{g}$.

One assumes the ratio between the energy differences related to $\alpha$-relaxation $E^{(\alpha)}$ and the range of glass transition temperatures of the elements is derived from the heat capacity gap near the glass transition temperature as
\begin{equation}
\delta E^{(\alpha)} = c_V^{(\alpha)} \delta T_g^{(\alpha)}
\label{eqDeltaE}
\end{equation}
where $c_V^{(\alpha)}$ is the isochoric heat capacity gap near the glass transition temperature of the volume element and the $\delta E^{(\alpha)}$ and $\delta T_g^{(\alpha)}$ are taken as fluctuations in $E^{(\alpha)}$ and $T_g$ between different elements.
\stmod{One then assumes that these energy fluctuations between the volume elements are related to thermodynamic fluctuations, and so are related to the heat capacity:}
\begin{equation}
(\delta E^{(\alpha)})^2 = \langle (\delta E^{(\alpha)})^2 \rangle = k_B T^2 c_V^{(\alpha)}.
\label{eqDeltaE2}
\end{equation}
Now the size of the volume elements enables us to infer a lengthscale $\xi=[V^{(\alpha)}]^{1/3}$ by relating the \emph{reciprocal} heat capacity gap of each element to the reciprocal isochoric \emph{molar} heat capacity gap $\Delta\left( 1/C_V\right)$\cite{donth1982, donth1996, hempel2000}. By combining Eqs. \ref{eqDeltaE} and \ref{eqDeltaE2} with $\Delta C_V=c_V^{(\alpha)}/N_A\rho V^{(\alpha)}$, where $N_A$ and $\rho$ are Avogadro's number and the number density of dimers, respectively.
\begin{equation}
V^{(\alpha)} = \xi^3 = \Delta \left( \frac{1}{C_V} \right)  \frac{R \langle T_g \rangle^2}{\rho (\delta T_g)^2}
\label{eqDeltaE3}
\end{equation}
where $R=N_Ak_B$ is the gas constant. In order to obtain a measure for $\Delta(1/C_V)$, we approximate with the molar isobaric heat capacity $C_p$ as, $\Delta\left(1/C_V\right) \approx 0.74\Delta\left(1/C_p\right)$ \cite{hempel2000, zheng2008}.

The resulting lengthscale $\xi$ is closely related to cooperatively rearranging regions (CRR) of Adam-Gibbs or random first order transition theory. This may be obtained from a thermodynamical treatment \cite{yamamuro1998, yamamuro1999, tatsumi2012}. Here we use the molar isobaric heat capacity per dimer $C_p$.

\subsection{Change in co-operative lengthscale with composition}
\label{sectionChange}

\begin{figure}
\begin{center}
\includegraphics[width=0.45\textwidth]{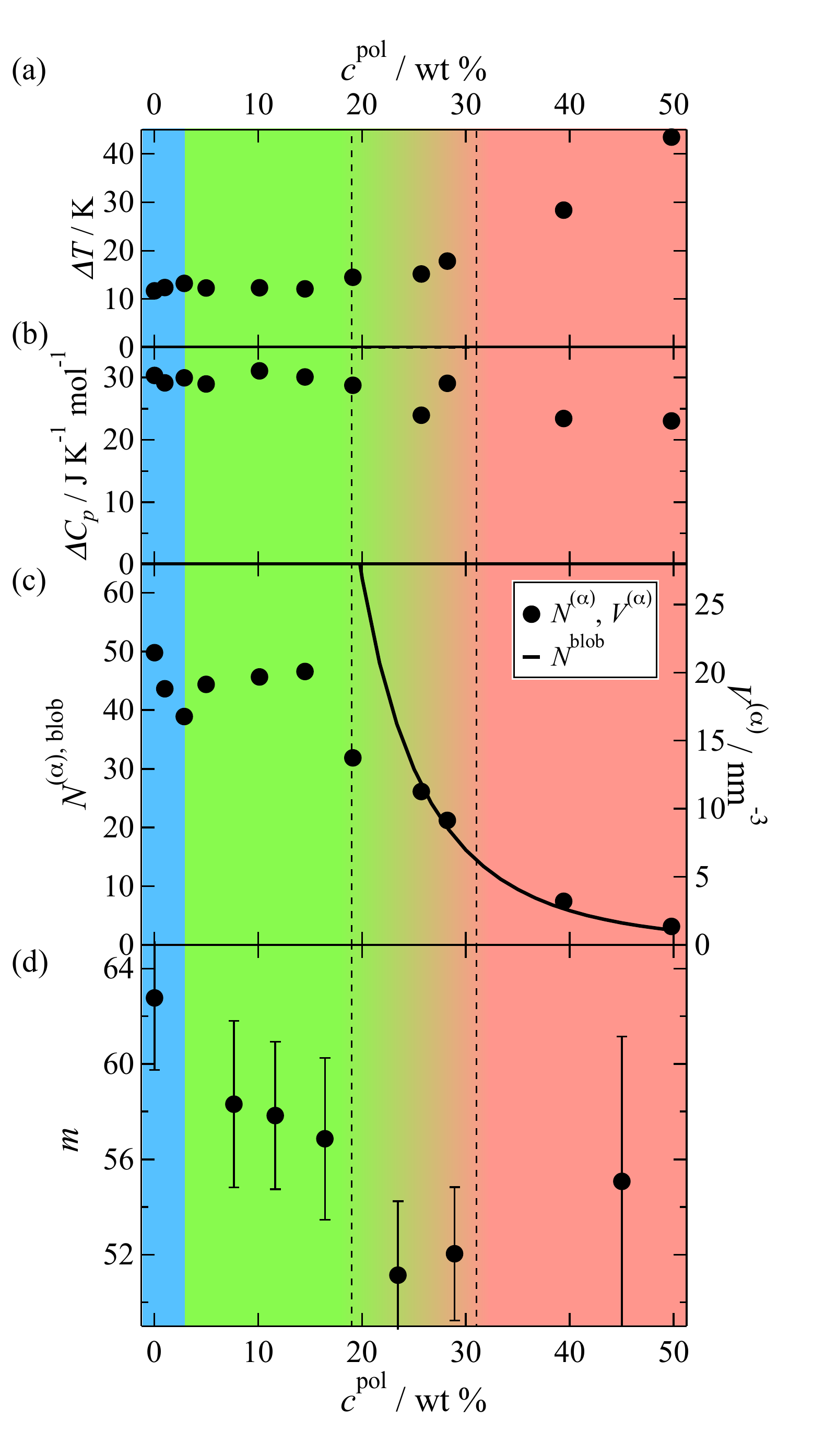}
\caption{(color online) $c^\mathrm{pol}$ dependences of 
(a) The temperature width of the glass transition $\Delta T$, 
(b) Molar isobaric heat capacity gap $\Delta C_p$, and 
(c) Number of dimers $N^\mathrm{(\alpha)}$ and volume $V^\mathrm{(\alpha)}$ associated with relaxation respectively. 
$N^\mathrm{blob}$ is the number of dimers in a blob. 
(d) Calculated fragility \paddyspeaks{parameter $m$} (Eq. \ref{eqFragility}) obtained by dielectric relaxation experiments.
In all panels, broken lines are set around 20 and 30 wt\% for the guide to eyes.}
\label{figDonth} 
\end{center}
\end{figure}

The results shown in Fig.~\ref{figDonth} clearly demonstrate the effect of polymer concentration in different regimes --- semidilute, with polymer weight fraction less than $\sim$20--30\% --- and the concentrated regime. Upon increasing polymer weight fraction,  $T_{\mathrm g}$ and $\Delta T$ only slightly increase in the semidilute regime, but rapidly increase at concentrated regime, whereas the heat capacity change $\Delta C_p$ is almost constant in each region with just a small drop from the semidilute to the concentrated regimes.

\paddyspeaks{In the context of our assumptions based on the Donth theory (preceding section),} 
these results \paddyspeaks{imply} 
a significant difference in the size of cooperative regions $N^{(\alpha)}$ and volume $V^{(\alpha)}$ between the semidilute and concentrated regimes. In the semidilute regime, $N^{(\alpha)}$ and $V^{(\alpha)}$ are almost constant with $N^{(\alpha)}\approx40-50$ (or $V^{(\alpha)}\approx20$ $\mathrm{nm}^{3}$) in the concentrated regime. However, in the concentrated regime, $N^{(\alpha)}$ and $V^{(\alpha)}$ drop significantly to around $N^{(\alpha)}\approx5$ and $V^{(\alpha)}\approx2$ $\mathrm{nm}^{3}$. That is to say, there is very little cooperativity. In the semidilute regime, the degree of cooperativity is consistent with previous studies corresponding to a fragile glass former \cite{hempel2000, zheng2008}, including the pure hexamer or pentamer \cite{zheng2008} while the very small cooperativity in the concentrated regime is \paddyspeaks{suggestive} 
of a strong glass former which would have no ideal glass transition \cite{chakrabarty2015}.

Now the size of co-operatively re-arranging regions determined following Adam-Gibbs theory is substantially smaller than the size of cooperativity following Donth's theory \cite{sillescu1999}. While we emphasise that determination of these lengthscales is very challenging and is \emph{indirectly inferred}  \cite{ediger2000,royall2015physrep,karmakar2014}, one possible explanation for the discrepancy between assumptions based on Donth or Adam-Gibbs theory is that once the relaxation has occurred in a  co-operatively re-arranging region, some surrounding molecules may move to some extent, to which some approaches may be more sensitive 
\paddyspeaks{to this motion} than others. Indeed, the lengthscale of dynamical heterogeneity estimated by 4D-NMR experiments at a temperature of $T_\mathrm{g} + 10 K$ \cite{tracht1998, tracht1999, ediger2000} is about 2-4 nm which is similar to our results in the semidilute regime. Other experimental methods to measure dynamical heterogeneity include forward recoil spectrometry \cite{swallen2003} and measuring oxygen diffusion \cite{syutkin2010}, both of which also give values consistent with the data quoted here. \stcomt{Soichi: Need to refer Smarajit paper here?}

From a microscopic point of view, it is interesting to compare the free volume of regions between polymers. In the semidilute regime, the blob model describes the whole system as a collection of tight packing blobs constituted from only one polymer chain \cite{teraoka}. Here the blob size $\xi^\mathrm{blob}$ is 
\begin{equation}
\xi^\mathrm{blob} = R_g\left( \frac{c^\mathrm{pol}}{c^\mathrm{pol}_*}\right)^{-\nu}
\label{eqBlobsize}
\end{equation}
where, $c^\mathrm{pol}_* \approx 4.04 \mathrm{wt\%}$ is the lower boundary of the semidilute regime, $\nu=1$ is the exponent which relates to molecular weight dependence of the polymer radius of gyration, where $R_g$ is the radius of gyration in the dilute limit. For poly $\alpha$-methyl styrene we have $R_g \approx 15.9$ nm \cite{osa2000}. Since $\xi^\mathrm{blob}$ reflects the typical interchain distance, the number of dimers in that space is estimated as \cite{teraoka},
\begin{eqnarray}
N^\mathrm{blob} &=& (1-c^\mathrm{pol})\frac{\rho N_A}{m_\mathrm{w}^\mathrm{d}} (\xi^\mathrm{blob})^3
\end{eqnarray}
where $m_\mathrm{w}^\mathrm{d}$ is molecular weight of the dimer and $m_\mathrm{w}^\mathrm{d}$ is 236.35. $N^\mathrm{blob}$ denotes the estimated number of dimers. 
We plot $N^\mathrm{blob}$ as a function of composition in Fig.~\ref{figDonth}, and we see that for low polymer concentration $c^\mathrm{pol} \lesssim 20\mathrm{wt\%}$, the number of dimers per blob is much larger than $N^{(\alpha)}$ where is almost constant at around 40--50. For higher concentrations, the number approaches the value of $N^{(\alpha)}$, which means that constraints from polymer chain apparently affect the number of cooperative molecules. 

\subsection{Dielectric relaxation measurements}
\label{sectionDielectric}

\begin{figure*}
\begin{center}
\includegraphics[width=\textwidth]{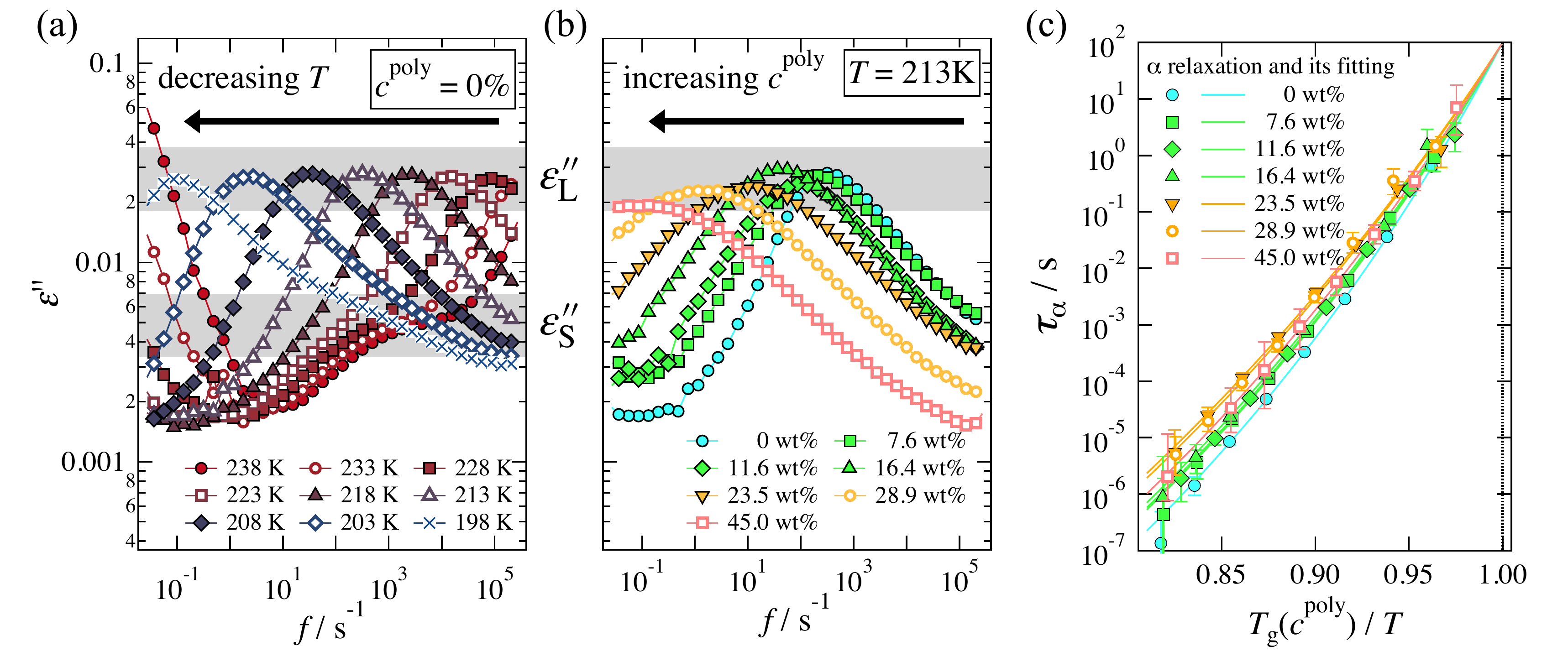}
\caption{(color online) The variations of the imaginary part of measured capacitance as a function of frequency at (a) different temperatures (198 K $\leq T\leq$ 238 K) for pure dimer and (b) different polymer concentration (0\% $\leq c^{\mathrm{poly}} \leq$ 45\%) for those at 213 K. \stmod{Gray shaded regions in Fig.(a,b) indicates corresponding signals from large ($\mathrm{L}$) and small ($\mathrm{S}$) \paddyspeaks{contributions (Eq. \ref{eqHN}).}} Figure (c) indicates the obtained relaxation time in time domain as a function of inverse temperature normalized with glass transition temperature. Lines indicate the Vogel-Fulcher-Tamman fitting results for each samples. Details were shown in the main text. 
}
\label{figDRSummary} 
\end{center}
\end{figure*}

\textit{Data collection. --- }
Figure \ref{figDRSummary}(a-b) shows the variation of the imaginary ($\varepsilon''$) part of the measured \paddyspeaks{dielectric} permittivity of the system as a function of frequency (0.03 Hz $\leq f \leq$ 3.0$\times 10^5$Hz) at different temperatures (198 K $\leq T\leq$ 238 K) for the pure dimer (Fig.\ref{figDRSummary}(a)) and for different polymer concentrations (0\% $\leq c^\mathrm{pol} \leq$ 45\%) for samples at 213 K. Peaks of the imaginary part of the permittivity are clearly seen in these results. In Fig.\ref{figDRSummary}(a), those peaks are understood as the superposition of a large $(\stmod{\mathrm{L}})$ signal
whose peak value is around 0.03 and smaller $(\stmod{\mathrm{S}})$ signal whose peak value is around 0.004. This spectroscopic data can be
understood in terms of the superposition of two Havriliak-Negami (HN) contributions $(\stmod{\mathrm{L}})$ and $(\stmod{\mathrm{S}})$, 
\begin{eqnarray}
\varepsilon^*(\omega) = \varepsilon_\infty + \frac{\Delta \varepsilon_\stmod{\mathrm{L}} }{\left[1+\left(i \omega \tau_\stmod{\mathrm{L}}\right)^{\alpha_\stmod{\mathrm{L}}}\right]^{\gamma_\stmod{\mathrm{L}}}}+ \frac{\Delta \varepsilon_\stmod{\mathrm{S}} }{\left[1+\left(i \omega \tau_\stmod{\mathrm{S}}\right)^{\alpha_\stmod{\mathrm{S}}}\right]^{\gamma_\stmod{\mathrm{S}}}}
\label{eqHN} 
\end{eqnarray} 
where, $\omega$, $\Delta \varepsilon_{\stmod{\mathrm{L,S}}}$, $\tau_{\stmod{\mathrm{L,S}}}$, $\alpha_{\stmod{\mathrm{L,S}}}$, and $\gamma_{\stmod{\mathrm{L,S}}}$ denote angular frequency $2 \pi f$, magnitude of \stmod{dielectric} permitivity loss, typical relaxation time, and nonlinear parameters in HN equation, respectively. Throughout our analysis, large permitivity loss, where $\Delta \varepsilon_{\stmod{\mathrm{L}}}\simeq $0.03 in Fig.\ref{figDRSummary}(a), is assigned to $\alpha$-relaxation which leads to glass transition. Another small dielectric loss is considered as intramolecular rotational freedom in dimer and is not central to our interests here.

\textit{Conversion to time-domain relaxation. --- }
As a rough approximation, one can treat the Havriliak-Negami (HN) equation \ref{eqHN}  as a representation of a stretched exponential relaxation function 
($\phi(t)=\exp\left[-(t/\tau_\stmod{\rm \alpha})^\beta \right]$) 
in the frequency domain, where $\tau_\stmod{\rm \alpha}$ and $\beta$ are the relaxation time and non-linear parameter in \stmod{the} time domain function, respectively.
However it is desirable to convert the results of the HN equation \paddyspeaks{from the frequency domain} into the time domain. Here we have elucidated the time domain relaxation function based on the assumption of a stretched exponential form. Work by Alvarez {\it et al.}\cite{alvarez1991, alvarez1993} demonstrated the coincidence of the HN equation and stretched exponential equation with a wide range of parameter space. \paddyspeaks{While it is clear that pinning some fraction of the system would not necessary result in a decay of the ISF to zero, we nevertheless use this parameterisation to estimate the} 
relaxation time $\tau$ as a function of temperature and polymer concentration, respectively.


The \paddyspeaks{structural} relaxation times thus obtained 
$\tau$  
are plotted as a function of inverse temperature normalized with $T_\mathrm{g}$ in Fig.\ref{figDRSummary}(c) and fitted with the  Vogel-Fulcher-Tammann (VFT) equation, 
\begin{eqnarray}
\tau_\mathrm{vft} = \tau_0 \exp\left(\frac{DT_0}{T-T_0}\right).
\label{eqVFT}
\end{eqnarray}
\paddyspeaks{where
  \stmod{$\tau_0$, $D$, and $T_0$}
   are fitting parameters.} The fitting parameters for each polymer concentration are listed in Tab.\ref{tabFITLIST}. Here we assumed the system has a single $\tau_0 $ for all $c^{\mathrm{poly}}$, as all plots refer to the same dimer. Finally the resulting fragility index, $m$, 
\begin{equation}
m(c^{\mathrm{poly}})=\frac{d [\log \tau_\mathrm{vft}(T,c^{\mathrm{poly}})]}{d(1/T)}\biggr\rvert_{T=T_g(c^{\mathrm{poly}})}
\label{eqFragility}
\end{equation}
is plotted as a function of polymer concentration in Fig. \ref{figDonth}(d). \paddyspeaks{Here we determine $m(c^{\mathrm{poly}})$ from the VFT equation \ref{eqVFT}.
There is some evidence for a weak drop in the fragility upon pinning, from the values of the fragility parameter $m$ plotted in Fig. \ref{figDonth}. However, within our error bars, 
it is hard to be sure. 
Visual inspection of the Angell plots in Fig. \ref{figDRSummary}(c) similarly suggests a small decrease in fragility with increasing polymer concentration.}

Consistent with results from the scanning calorimetry, while the lower the polymer concentration the more fragile the system is, the higher the polymer concentration, namely larger than 20 wt\%, the stronger the system is. This relation is easily seen from the Fig.\ref{figDRSummary}(c), here the lower the polymer concentration, the steeper the slope at $T_\mathrm{g}$. However, it is worth noting that the changes in fragility here are modest, and less than might be expected from the change in the size of cooperative regions that we have inferred from the DSC data.

\begin{table*}[t]
\begin{threeparttable}
\caption{List of fitting parameters obtained by VFT equation in each polymer concentration. }\label{tabFITLIST}
\begin{tabular}{c|p{3cm}p{3cm}p{3cm}p{3cm}p{3cm}}
\hline
$c_\mathrm{pol}$[wt\%] & $\log \left(\tau_0/s\right)$ \tnote{a}& $T_\mathrm{g} / K$ & $m$ & $T_0 /K $ & $D$ \\
\hline 
$0$ & $-19.0\pm1.6$ & $190.47\pm0.43$ & $62.8\pm3.0$ & $126.9\pm5.7$ &$24.2\pm4.9$ \\
$7.6$ & $-19.0\pm1.6$ & $190.80\pm0.82$ & $58.3\pm3.5$ & $122.2\pm6.6$ &$27.1\pm5.8$ \\
$11.6$ & $-19.0\pm1.6$ & $192.95\pm0.67$ & $57.8\pm3.1$ & $123.0\pm6.5$ &$27.4\pm5.8$ \\
$16.4$ & $-19.0\pm1.6$ & $194.82\pm0.81$ & $56.9\pm3.4$ & $123.0\pm6.9$ &$28.2\pm6.1$ \\
\hline
$23.5$ & $-19.0\pm1.6$ & $196.31\pm0.95$ & $51.1\pm3.1$ & $115.8\pm7.8$ &$33.5\pm7.6$ \\
$28.9$ & $-19.0\pm1.6$ & $200.57\pm0.74$ & $52.0\pm2.8$ & $119.8\pm7.5$ &$32.6\pm7.2$ \\
$45.0$ & $-19.0\pm1.6$ & $207.7\pm1.5$ & $55.1\pm6.1$ & $129\pm11$ &$29.7\pm7.9$ \\
\hline
\end{tabular}
\begin{tablenotes}[para,flushleft,online,normal] 
\item[a] The parameter $\tau_0$ is fixed throughout the entire analysis.
\end{tablenotes}
\end{threeparttable}
\end{table*}

\section{Discussion and Conclusions}
\label{sectionDiscussion}

Here we demonstrate a means to suppress molecular relaxation in a manner inspired by pinning by using a dimer-polymer mixture. \paddyspeaks{Conceptually, we used the polymer to approximate a ``pinning field''.} We expect that realization of the transition to the ideal glass is related to the concentration of pinning particles, and that sufficiently high concentration would destroy the transition such that the relaxation time of the system would simply grow without any thermodynamic transition. In fact, Cammarota and Biroli \cite{cammarota2012pnas} demonstrated the existence of a critical point in the temperature -- pinning concentration plane. In the case of the renormalization group approach there is a critical point with a fraction of pinned particles of about 0.22 below which the transition to the ideal glass is expected whereas above this concentration of pins, no transition to ideal glass is expected. 
\paddyspeaks{Some hint of this is seen in our results, which are compatible with a drop in fragility in the case of strong ``pinning''.}

Combining \stmod{differential}
 scanning calorimetry and dynamic dielectric 
\stmod{spectroscopy} 
\paddyspeaks{gives a two--pronged approach to 
\stmod{investigate }
both thermodynamic and dynamic aspects of the system.}
In particular, for weaker pinning ($\lesssim$20 wt\%), while we find a constant co-operativity length from thermodynamical measurements, the fragility of those system, obtained by dielectric relaxation, is about 60, the system is fragile. On the other hand, in the stronger pinning regime ($\gtrsim$20 wt\%), while we find a drop in the co-operativity length with increasing polymer concentration from thermodynamical measurements, the fragility as deterimed by by dielectric relaxation shows a modest drop to around 50. This behaviour is \paddyspeaks{somewhat} reminiscent of previous computer simulations \cite{chakrabarty2015}. 

Here we have restricted the interactions between the molecules and polymer to be identical by taking, dimers of the same monomer. However, as reported in previous studies, the polymer concentration dependence of $T_\mathrm{g}$ does not agree with traditional theory \cite{scandola1982, pizzoli1983, scandola1987}. The behavior of molecules in a semidilute polymer solution may therefore be understood within the context of pinning. Moreover, taking the viewpoint that pinning is equivalent to certain constraints, there is a possibility that previous studies in confined systems such as zeolites or silica \cite{oconnell1999, nagoe2010, nagoe2015jpcm}, all of which show a broadening of the glass transition, could be understood with the context of pinning.

Along with other approaches, which use larger molecules as ``soft pins'' \cite{das2017,das2021}, our work opens up the question of how the accessible routes towards pinning in experimental molecular systems influence the nature of the glass transition. The effects of polymerisation of the pins (rather than random pinning) and the non-equilibrium configuration of the vitrified polymer pins are important quantities to investigate theoretically or computationally, to enable this route towards an ideal glass to be placed on a firmer footing. \paddyspeaks{Our work may also be extended to other systems and indeed raises the question of whether a development of the work by Alvarez {\it et al.}\cite{alvarez1991, alvarez1993} might be extended to consider the case where the ISF does not fully decay.}

\begin{acknowledgements}
We thank Smarajit Karmakar for useful discussions.
ST would like to thank with Prof. Susumu Fujiwara and Prof. Masato Hashimoto for letting us use their DSC equipment. 
ST would like to thank with "Top Global University Japan Project", MEXT, Japan and the Promotion of Science (JSPS), which is supported in Kyoto Institute of Technology. ST was dispatched to University of Bristol as a visiting fellow for a duration of 1 year in 2016.
CPR acknowledges the Royal Society and European Research Council (ERC Consolidator Grant NANOPRS, project number 617266).
\end{acknowledgements}
\bibliographystyle{apsrev}
\bibliography{lalaland}

\end{document}